\documentstyle[12pt,aaspp4,psfig]{article}

\lefthead{Deltorn et al.}
\righthead{A massive cluster at $z\sim1$}

\begin{document}

\title{A MASSIVE CLUSTER OF GALAXIES AT z = 0.996}

\author{J.-M. Deltorn\altaffilmark{1}, O. Le F\`evre\altaffilmark{1}}
\affil{DAEC, Observatoire de Paris-Meudon, 92195 Meudon Cedex, France}

\author{David Crampton}
\affil{Dominion Astrophysical Observatory, National Research Council of
Canada, R.R. 5 Victoria, B.C., V8X4M6, Canada}

\and

\author{M. Dickinson}
\affil{Space Telescope Science Institute, Baltimore, MD 21218, USA}

\altaffiltext{1}{Visiting Astronomer, Canada France Hawaii Telescope. 
CFHT is operated by the National Research Council of Canada, the Centre
National de la Recherche Scientifique of France, and the University of
Hawaii.} 

\begin{abstract}
We report the identification of a cluster of galaxies around the
high-redshift radio galaxy 3CR184 at z = 0.996. The identification is
supported by an excess of galaxies observed in projection in
$I$ band images (both in ground-based and HST data), a peak in the redshift
distribution comprising 11 galaxies (out of 56 with
measured redshifts) in a $\sim2000$ km s$^{-1}$ velocity interval, and the
observation on HST WFPC2 frames of a gravitational arc seen projected at
$42h_{50}^{-1}$kpc away from the central radio galaxy. We thus have
strong evidence for the presence of a massive cluster at
$z\simeq1$.  The mass contained within the arc radius is in the range $[1.20\times10^{13}h_{50}^{-1}M_{\odot},2.78\times10^{13}h_{50}^{-1}M_{\odot}]$ for $z_{arc}$ within the interval 3-1.5; the corresponding mass to light ratio varies from $56h_{50}$ to $140h_{50}$. The velocity dispersion deduced from the galaxy cluster redshifts is ${634}^{+206}_{-102}$km s$^{-1}$, leading to a virial mass $M=6.16_{-2.40}^{+3.94} \times 10^{14} h_{50}^{-1}M_{\odot}$ and a mass to light ratio $200h_{50}<(M/L_B)_{400h_{50}^{-1}}<500h_{50}$ within a radius of $400h_{50}^{-1}$kpc.

\end{abstract}


\section{INTRODUCTION}

Understanding the formation and evolution of large scale structures is of
considerable importance to modern cosmology. While galaxy formation and
evolution is receiving much warranted attention, the observational study of
the formation and evolution of clusters of galaxies or other large scale
structures is only now developing. 

To identify  clusters at high redshifts is a challenging observational
task.  At low redshifts, one can rely  on the projected 2D galaxy
density, as was done successfuly from the earliest photographic
material (e.g., Abell 1958, Zwicky 1961-1968).  However, at high redshifts,
$z\geq0.5$, it becomes increasingly difficult to identify the 2D galaxy
overdensity produced by a cluster because the projected foreground and
background galaxies are so numerous that the density contrast is
severely reduced and chance alignments of groups of galaxies can
mimic the appearance of clusters (Frenk et al. 1990).

The existence of a distant cluster can be proven only if several lines of
evidence are combined, which should include the measurement of a
projected overdensity of galaxies in few
square arcmin area on the sky, combined with a  significant overdensity
observed in the redshift distribution of galaxies on velocity scales
$\sim$2000 km s$^{-1}$ in order to reduce the contamination by
foreground and background interloper galaxies.  In addition, evidence
for hot gas, through the detection of X-ray emission, and/or evidence
for a peaked mass distribution as indicated by weak or strong lensing
of background galaxies, provide strong support for the identification
of such overdensities as genuine clusters of galaxies.  Although a
number of candidates have been proposed, only a few clusters (or
proto-clusters) of galaxies have been unambiguously identified at z$
>0.6$, with the above criteria fulfilled (Dickinson 1996; Luppino
\& Kaiser 1996). It is of considerable importance for our knowledge of
the evolution of large scale structure to identify more high redshift
clusters in order to establish the evolution of their physical
properties with redshift.

One possible search strategy is to look for clusters of galaxies
around known powerful radio galaxies (Yee \& Green 1984; Hill \& Lilly
1991; Dickinson 1996; Le F\`evre et al. 1996). We present here the
unambiguous identification of a cluster of galaxies around the
radio galaxy 3CR184, at a redshift z = 0.996.

H$_0=50$ km s$^{-1}$Mpc$^{-1}$ and $q_0=0.5$ are used throughout this
{\it Letter}.

\section{CFHT IMAGING AND SPECTROSCOPY}

The field around 3CR184 was imaged in 1994 January with the
Multi-Object-Spectrograph (MOS) at CFHT  (Le F\`evre et al 1994) in
imaging mode. A 900s image was obtained in the $R$ band and deep images
were also obtained in a narrow band filter centered on 7480\AA\  with a
bandwidth of 200\AA, which includes the [OII]3727\AA ~line redshifted
to z$\sim1$, for a total integration time of 3600s. Additional $I$ band
images were obtained with the same instrumental setup on 1994 February
6 for a total integration time of 2400s.
An initial selection of spectroscopic targets was performed on the
basis of  excess emission in the narrow band images. Spectroscopic
follow-up was performed on 1994 February 6 with MOS in its multi-slit
mode. A multi-slit mask with 37 slits, 2 arcseconds in width and at
least 10 arcseconds in length each, was used to obtain three
spectroscopic exposures of 4800s each. The R300 grism provided a
spectral resolution of 23\AA, and a wavelength coverage from 5000\AA\
to 1$\mu$m.  Data reduction was performed with the {\sc MULTIRED}
software implemented under {\sc IRAF} (Le F\`evre et al. 1995). Two
galaxies were subsequently identified at a redshift within a few
hundred km s$^{-1}$ of the radio galaxy, indicating the possible
presence of a cluster.
Additional spectra were obtained during 1995 December 20-24 with the
same instrumental setup as described above. The object selection this time was
based purely on the $I$-band magnitudes and spatial location on the
plane of the sky. Three additional multi-slit masks were used to obtain
spectra for 122 objects. Integration times of 15000, 14400, 15000s were
obtained for masks 2, 3, and 4, respectively. The linear geometry of
the slit positions on each mask result in a selective sampling of
galaxies on the sky. All of the objects observed are in an E-W strip of
8\farcm5 $\times$ 2\farcm85 and the ratio of spectroscopically
identified objects to the total number of sampled objects fainter than
the radio galaxy ($I_{RG} = 19.65$), and brighter than $I<22.2$, is 0.35
in this strip.

Data processing followed the procedure outlined in Le F\`evre et al.
(1995). Of 122 objects observed, 56 were identified as galaxies, 26
turned out to be galactic stars, 1 was identified as a quasar and
the remaining 39 were unidentified. A more detailed description of the
observational data  will be given elsewhere (Deltorn et al. 1997).

\section{HST IMAGING}

HST imaging was conducted with the Wide Field and Planetary Camera 2 
during HST cycle 5. A total exposure time of  11000s was
obtained with the F814W filter and 6600s with the F606W filter.
Exposures were shifted by integral pixel values to allow for cosmic rays
and bad pixel cleaning. Standard HST pipeline data reduction was performed.
Photometry of 878 objects in the field was obtained, based on the HST
photometric zero point calibration converted to Vega units using the color corrections of Holtzman et al. (1995). The completeness limit is $I=26$,
while objects can be detected down to $I=28$.
The image of a region $24\times24$ arcsec$^2$ around the radio galaxy is shown in
Figure 1 (Plate 1). 
Numerous faint galaxies can be seen in the immediate
surroundings of the radio galaxy. The radio galaxy and two galaxies with
spectroscopically confirmed redshifts which indicate membership in the cluster were imaged in the WFPC2 field; their properties will be discussed
elsewhere (Deltorn et al. 1997).

One of the most interesting features in the images is the faint
arc-like structure located 4\farcs9 to the north-east of the radio galaxy. 
The arc is $\sim$ 3\farcs6 long and is visible in both the F606W and 
the F814W images.
Although it is possible that it is due to a chance superposition of
several faint galaxies, the center of curvature of the arc is coincident with the
radio galaxy and it is unresolved in the radial dimension. The arc has a
secure detection $S/N=2.8$ in the F814W image and $S/N=3.4$ in the F606W
filter. The magnitude of the arc is $I=25.0\pm0.4$ and $V-I=0.3\pm 0.8$.
 The presence
of this gravitational arc close to the central radio galaxy indicates
the proximity of a high concentration of mass.

\section{EVIDENCE FOR CLUSTERING AROUND 3CR184 AT $z\sim0.996$}

In order to quantify any projected excess number of galaxies in the
vicinity of 3CR184 we have computed density maps using the 
estimator $D_{proj}$ defined in Dressler (1980).
 Cuts in magnitude
were applied on the data in order to consider only objects that
are fainter than the radio galaxy and brighter than $I \leq 22.5$ for
ground-based images and fainter than the radio galaxy and
brighter than $I \leq 26.5$ for
HST data, the latter limiting magnitude corresponding to $M_B\simeq-17$ at rest, for objects at the redshift of 3CR184. Color selection was also introduced; since high redshift
early-type galaxies are on average redder than lower redshift ones we
retained only those galaxies with $R-I \geq 0.2$ for ground-based data
and with $V-I \geq 1$ for the HST data. Those objects should
predominantly lie at redshifts greater than 0.5, increasing the
projected galaxy density contrast of high redshift structures.

The density maps constructed from the magnitude-limited and
the color-limited samples for the HST data show a peak excess density around the radio galaxy
corresponding to a $5\sigma_{bg}$ and $10\sigma_{bg}$ excess above the mean
galaxy background respectively; where $\sigma_{bg}$ corresponds to the square root of the variance of $D_{proj}$ measured outside a radius of 1\arcmin$\,$ centered on the radio galaxy. The density peak lies 6\arcsec$\,$  east of the radio
galaxy. The ground-based images show lower excess of $4\sigma_{bg}$ and $7\sigma_{bg}$ around 3CR184 for the magnitude-limited and
the color-limited samples respectively.
The central richness, $N_{0.5}^c$ (Bahcall 1981), was computed
after correction for the mean galaxy background  estimated
 using the deep  galaxy counts of Abraham
et al. (1996) in the Hubble Deep Field, which
led to a central richness of $N_{0.5}^c\simeq 39 \pm 9$, the error being
calculated assuming purely Poisson statistics. This excess
corresponds roughly to an Abell richness class 2 cluster. The
significance of the measured excess - far above the statistical
``noise'' - and the presence of redder objects around the radio galaxy
(the excess increasing with the V-I cut), combine to give confidence in
the reality of the projected 2D overdensity.
Figure 2 shows the redshift distribution of the galaxies in our sample.
From a total of 56 securely identified objects, 11 galaxies, including
the radio galaxy, have velocities within +1198/$-$750 kms$^{-1}$ of
that of 3CR184. The clear peak at $z\simeq1$ apparent in Figure 2
demonstrates the presence of an overdensity in velocity space around
3CR184. The list of galaxies within this peak is given
in Table 1.  The redshift distribution expected for a similar size
sample of field galaxies, as measured from deep redshift surveys
(Crampton et al. 1995), indicates that 0.75 galaxies should be observed
in a random sample of field galaxies in the same velocity bin, down to $I\simeq22$. It is
therefore highly improbable that this excess is due to a random
distribution.

The observation of an excess density both in the projected
$(\alpha,\delta)$ space and in redshift space demonstrates
unambiguously the reality of the clustering of galaxies around the
bright radio galaxy 3CR184. This, combined with the high concentration
of mass, as demonstrated by presence of a gravitational arc, secures the
identification of a cluster around 3CR184. Future deep X-ray observations of 
this field would provide useful complementary informations about the
hot gaz emission of the cluster.

\section{VIRIAL AND LENSING MASS ESTIMATES}

The observation of a gravitational arc associated with a cluster of
galaxies
allows a direct determination of the mass within the perimeter defined
by the arc. If a spherically-symmetric projected mass 
distribution is assumed, and the lensed galaxy is coincident 
with the direction of the cluster center, we have
$M_{proj}(\theta_{arc})=0.25c^2G^{-1}\times D_{arc}D_{cl}D_{cl-arc}^{-1}{\theta}^2_{arc}$,
where $D_{arc}$, $D_{cl}$ and $D_{cl-arc}$ are the angular distances
from the lensed object, from the cluster, and between the cluster and
the lensed galaxy respectively, and $\theta_{arc}=r_{arc}/D_{cl}$.  We also assume $r_{arc}\simeq r_c$
($r_c$ being the Einstein radius) and the center of mass is taken to be
coincident with the radio galaxy.  For z$_{arc}$ in the range 1.5--3,
the mass enclosed within $42 h_{50}^{-1}kpc$ from the radio galaxy
varies from $2.78 \times 10^{13}$ to $1.20 \times 10^{13}h_{50}^{-1}M_{\odot}$,
with an average value of $1.68 \times 10^{13}h_{50}^{-1}M_{\odot}$.\\
If a singular isothermal sphere (SIS)
model is adopted, we then have the following central velocity
dispersion:
\begin{displaymath}
\sigma_{lens}=\left\{r_{arc}\frac{c^2}{4\pi}\frac{D_{arc}}{D_{cl}D_{cl-arc}}
\right\}^{\frac{1}{2}}.
\end{displaymath}
\noindent
This gives $\sigma_{lens} \in [990,650]$ km s$^{-1}$ for any value of
z$_{arc} \in [1.5,3]$ (the average value being 756 km s$^{-1}$),
assuming the arc is centered on the radio galaxy. In the framework of
SIS approximation, this velocity dispersion is independent of the
radius.

The measured redshifts of the cluster galaxies allow derivation
of a velocity dispersion in the framework of the cluster. 
We obtain $\sigma_{\|}={634}^{+206}_{-102}$km s$^{-1}$, for the sample 
of eleven galaxies which pass the exclusion criterion of Yahill \&
Vidal (1977).  The errors are estimated from both the redshift
uncertainties and the sampling errors due to the small number of
galaxies (Danese, De Zotti, di Tullio 1980). This is in good
agreement with the velocity dispersion as deduced from the lensing
configuration. 

Assuming the isotropy, the spherical symetry of the mass distribution and
the dynamical equilibrium of the structure we can derive an estimate of
the deprojected virial mass. In the case of equal masses of the galaxies, 
we found $M=6.16_{-2.40}^{+3.94} \times 10^{14} h_{50}^{-1}M_{\odot}$, the errors being
calculated from Danese et al. (1981).

The mass enclosed in a 1 $h_{50}^{-1}$Mpc radius can be estimated within the
isothermal sphere approximation. Considering a distribution
of galaxies with density profile $\rho \propto r^{-\epsilon}$, we
have:
\begin{displaymath}
M(<r)=\frac{\epsilon-2\beta}{\epsilon-\beta(\epsilon-1)}\frac{\epsilon\sigma_V ^2r}{G},
\end{displaymath}
where $\beta=0$ for isotropic orbits and $\epsilon\simeq2.2$ (Seldner $\&$ Peebles 1977). We find $M(\sigma_V=756$ km s$^{-1}, r=1h_{50}^{-1}Mpc)\simeq2.97 \times 10^{14}h_{50}^{-1}M_{\odot}$. Taking the velocity dispersion deduced from the cluster galaxy redshifts leads to a lower value of the
mass: $M(\sigma_V=634$ km s$^{-1}, r=1h_{50}^{-1}Mpc)\simeq2.06 \times 10^{14}h_{50}^{-1}M_{\odot}$.

The mass to light ratio can be calculated within the radius defined by the
arc. The total light within 4\farcs9  from the radio galaxy was corrected from
contamination due to field galaxies using the counts of Abraham et al. (1996) and the resulting I magnitude of the cluster population was then converted
to $M_B$, knowing that I band at a redshift of $\sim$ 1 roughly
corresponds to B band at rest, and assuming a no-evolution scenario for the spectral energy distribution of the cluster galaxies. We found a field-substracted luminosity of $1.16 \times 10^{12} h_{50}^{-2}L_{B_{\odot}}$, leading to $M/L_B \simeq 56h_{50}\,$within $42 h_{50}^{-1}kpc$. Using the
isothermal sphere approximation, we found, after correction for the field galaxies contamination, a total
luminosity within $400h_{50}^{-1}kpc$ of $L_{400h_{50}^{-1}}\simeq 2.30 \times 10^{12} h_{50}^{-2}L_{B_{\odot}}$ leading to  $(M/L_B)_{400h_{50}^{-1}}\simeq 200h_{50}$. In addition, if all cluster galaxies are subjected to a luminosity evolution similar to the one measured by the CFRS for field galaxies  in the redshift range $0.75<z<1$ (Lilly et al. 1995),the correction to the observed luminosity would give $M/L_B \simeq 140h_{50}\,$within $42 h_{50}^{-1}kpc$ and $(M/L_B)_{400h_{50}^{-1}}\simeq 500h_{50}$. We then believe that $200h_{50}<(M/L_B)_{400h_{50}^{-1}}<500h_{50}$. This range is comparable to the higher $M/L_V^{all}$ found by Smail et al. (1996) for their distant cluster sample ($z\in [0.17,0.56]$).\\

\section{DISCUSSION AND CONCLUSIONS}

  Our observations of the field around 3CR184 demonstrate unambiguously
the presence of a cluster of galaxies, with an associated mass
comparable to massive clusters observed locally.  We stress that the
identification is made possible by a combination of several
diagnostics:   excess in projected 2D density maps, excess in redshift space, 
lensing geometry, and mass estimate obtained from virial
analysis of spectroscopically measured galaxies.\\
The relative agreement,
 between the two estimates of the velocity dispersion - from the redshifts and the lensing analysis - converge in describing 
a quite massive cluster at the redshift of the radio galaxy.
We find a lensing mass of $1.68 \times 10^{13}h_{50}^{-1}M_{\odot}$ within 42h$_{50}^{-1}$kpc, 
and derive a virial mass of $6.16_{-2.40}^{+3.94} \times 10^{14} h_{50}^{-1}M_{\odot}$. 
The cluster around 3CR184 is thus one of the very few massive
structures yet identifed at redshifts larger than 0.9,
at epochs when the universe was less than 40\% 
of its present age.

While obtaining velocity dispersion and mass estimate is 
relatively straightforward, the description of the dynamical
state of the cluster is much more uncertain.
If a good agreement between virial and lensing masses has been
demonstrated for some systems (e.g. PKS 0745-191: Allen, Fabian $\&$ 
Kneib 1996; MS2137-23: Mellier et al. 1993),
most arc-cluster associations reveal a significant discrepancy, 
likely due to the hypothesis
involving the mass/velocity dispersion definitions (Miralda-Escud\'e $\&$
Babul 1995; Wu 1996), indicating that these clusters may not be considered as
totally relaxed systems. 
In the case of the cluster around 3CR184, a comparision between the
lens and virial velocity dispersions is limited by the 
limited sample of cluster galaxies, the difficulty in
estimating the true center of mass of the presumed cluster,
and the simplistic assumptions necessary to derive either $M$ or
$\sigma$ (either an equilibrium state or a spherical symmetry).
 Given the current observational
dataset it is thus premature to speculate on the similarity between
$\sigma_{lens}$ and $\sigma_{\|}$ to draw any conclusion regarding the
dynamical state of the cluster around 3CR184.

The number of high redshift clusters seems to be steadily increasing 
as observational capabilities become more accute, and  
the  observation of this massive structure at $z\simeq1$,
combined with other observations of very high redshift clusters might 
come to be in severe
conflict with various cosmological models. Recently, the secure
identifications of massive bound structures at $z>0.8$, through
spectroscopic
observation of cluster members (Dickinson  1996; Deltorn et al.
1997, Francis et al. 1996), weak
gravitational lensing (Luppino $\&$ Kaiser 1996, Smail $\&$
Dickinson 1995), or X-ray 
observations (Luppino $\&$ Gioia 1995, Castander et al. 1994) have shown that those structures might not be as rare as 
predicted by some cosmological scenarios.
In CDM and even CHDM models, the number of high redshift massive clusters
predicted from both N-body simulations or the Press-Schechter formalism, seems to be too low with respect to
observations (Jing $\&$ Fang 1994). On the contrary, HDM models predict
too many high-z, high-$\sigma_v$ structures (Lijle 1990). Alternatives and modifications to
the
standard CDM scenario have been advanced in order to account for the emerging data on large scale structure and the amplitude of COBE fluctuations. Among
them, low $\Omega_0$ models (either flat or open) provide a significant
number density of massive high redshift clusters that may be compatible with
recent 
observations (Eke, Cole $\&$ Frenk 1996; Viana $\&$ Liddle 1996 and ref.
therein). As the abundance of massive structures provides increasing discrimination with increasing redshift, the identification, without
ambiguity,
of clusters at $z\simeq1$ is now beginning to provide useful observational
constraints
to the cosmological models.

\acknowledgments

We thank the CFHT director, P. Couturier for the allocation of
discretionary time to start this project, and the CFHT staff
for their support during the observations.

\figcaption[arc]{Sum of the HST F606W and F814W images of a $24\times24$ arcsec$^2$ field around 3CR184. The arc is located 4\farcs9 north-east of the radio galaxy and extends on 3\farcs6.}

\figcaption[fig_hist_z]{Redshift distribution of 56 galaxies in the field 
of 3CR184. The redshifts were obtained from MOS spectroscopy at CFHT. The bin
width corresponds to 2000 km s$^{-1}$ at the redshift of the radio galaxy (z=0.996).}

\clearpage

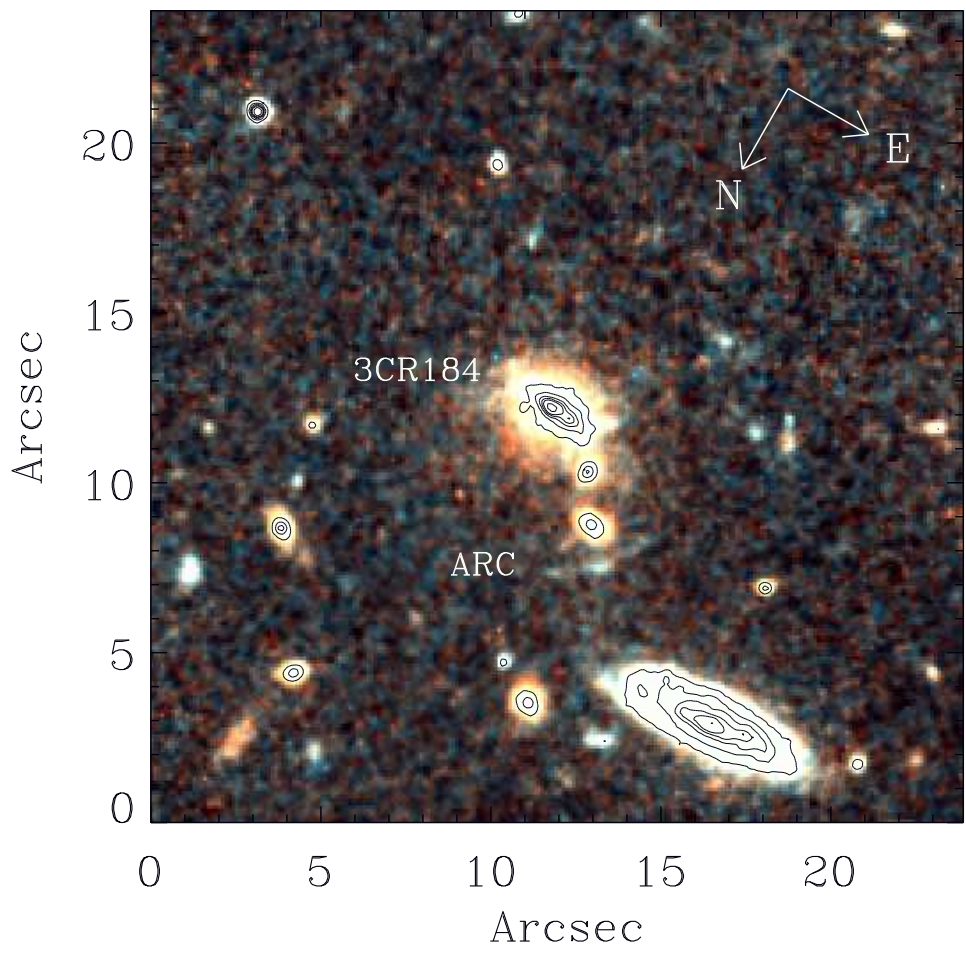
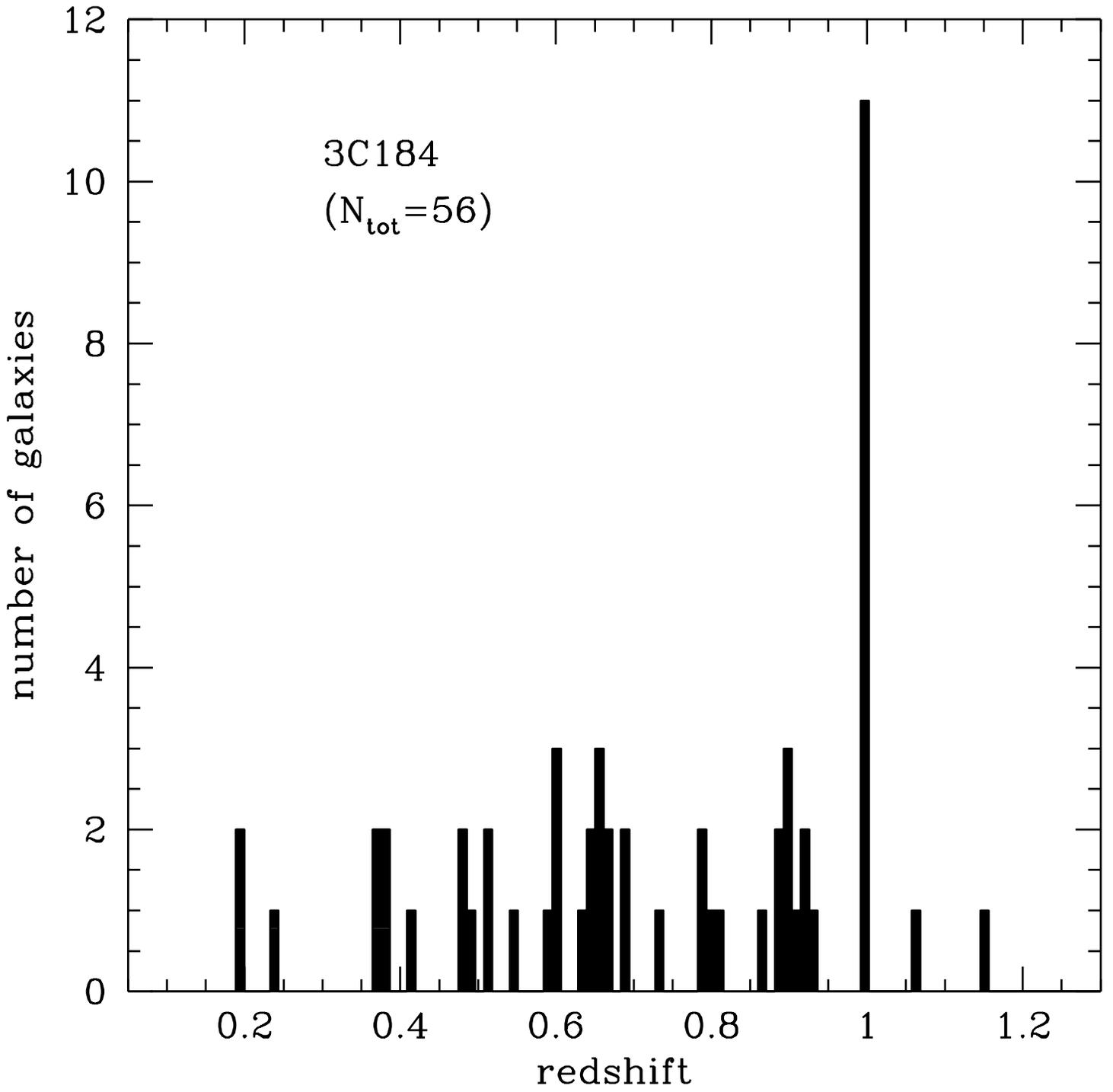
\end{document}